\documentclass[twocolumn,preprintnumbers,amsmath,amssymb,prl]{revtex4}

\usepackage{xspace}
\usepackage{graphicx}
\usepackage{dcolumn}
\usepackage{bm}
\usepackage{tabularx}
\usepackage{amsmath}
\usepackage{color}
\newcommand{\parallelsum}{\mathbin{\!/\mkern-5mu/\!}}

\begin{document}

\title{Unusual heat transport of the Kitaev material Na$_2$Co$_2$TeO$_6$: putative quantum spin liquid and low-energy spin excitations}
\author{Xi\v{a}och\'{e}n H\'{o}ng,$^{1,2,}\footnote{These authors contributed equally.}^{,}\footnote{x.c.hong@ifw-dresden.de}$ Matthias Gillig,$^{1,*}$ Richard Hentrich,$^{1,*}$ Weiliang Yao,$^3$ Vilmos Kocsis,$^1$ Arthur R. Witte,$^{1,4}$ Tino Schreiner,$^1$ Danny Baumann,$^1$ Nicol\'{a}s P\'{e}rez,$^1$ Anja U. B. Wolter,$^1$ Yuan Li,$^{3,5,}\footnote{yuan.li@pku.edu.cn}$ Bernd B\"{u}chner,$^{1,4,6}$ and Christian Hess$^{1,2,6,}\footnote{c.hess@uni-wuppertal.de}$}

\affiliation{$^1$Leibniz-Institute for Solid State and Materials Research (IFW-Dresden), 01069 Dresden, Germany\\
$^2$Fakult$\ddot{a}$t f\"{u}r Mathematik und Naturwissenschaften, Bergische Universit$\ddot{a}$t Wuppertal, 42097 Wuppertal, Germany\\
$^3$International Center for Quantum Materials, School of Physics, Peking University, 100871 Beijing, China\\
$^4$Institute of Solid State and Materials Physics and W\"{u}rzburg-Dresden Cluster of Excellence $ct.qmat$, Technische Universit$\ddot{a}$t Dresden, 01062 Dresden, Germany\\
$^5$Collaborative Innovation Center of Quantum Matter, 100871 Beijing, China\\
$^6$Center for Transport and Devices, Technische Universit$\ddot{a}$t Dresden, 01069 Dresden, Germany}

\date{\today}

\begin{abstract}
We studied the field dependent thermal conductivity ($\kappa$) of Na$_2$Co$_2$TeO$_6$, a compound considered as the manifestation of the Kitaev model based on the high-spin $d^7$ Co$^{2+}$ ions.
We found that in-plane magnetic fields beyond a critical value $B_c \approx$~10 T are able to drastically enhance $\kappa$ at low temperatures, resulting in a double-peak structure of $\kappa(T)$ that closely resembles the behavior of $\alpha$-RuCl$_3$. This result suggests that heat transport in Na$_2$Co$_2$TeO$_6$ is primarily phononic, and it is strongly affected by scattering from magnetic excitations that are highly tunable by external fields.
Interestingly, for magnetic fields $\textbf{B}\parallelsum \textbf{a}$ (i.e., along the zigzag direction of the Co-Co bonds), there is an extended field range which separates the long-range magnetic order for $B\leq B_c\approx10$~T and the partially spin-polarized gapped high-field phase for $B\gtrsim 12$~T.
The low-energy phonon scattering is particularly strong in this field range, consistent with the notion that the system becomes a quantum spin liquid with prominent spin fluctuations down to energies of no more than 2~meV.
\end{abstract}

\pacs{not needed}

\maketitle
A compass model of the effective spin$-1/2$ two-dimensional honeycomb lattice, known as the Kitaev model, is an emerging research field \cite{Kitaev,JvdB,Trebst,JPCM}.
Topological quantum spin liquids (QSL) with Majorana fermions as excitations are guaranteed as its ground states \cite{Kitaev}.
To accommodate pseudospin$-1/2$ magnetism and bond-dependent exchange couplings in real materials, heavy transition metal compounds in the low-spin $d^5$ electron configuration are preferred \cite{TakagiRev,JK}.
Guided by theoretical predictions \cite{JK,Rau,ChenG}, Ir$^{4+}$ oxides and Ru$^{3+}$ chloride have been studied extensively \cite{Takagi,Na213,a213,b213,g213,Cu213,Ag213,H213,CuLi213,CaoHB,Baenitz}.
Those studies imply that in real materials, the Heisenberg interactions ($J$) and off-diagonal terms ($\Gamma$) have to be also taken into account in the Hamiltonian.
Thus, instead of a pure Kitaev QSL, the ground states of these compounds are usually some form of magnetic order \cite{JPCM}.
Nevertheless, the Kitaev physics is still expected to be prominent in $proximate$ Kitaev materials as long as the Kitaev term ($K$) is relatively large \cite{Perkins,Pollmann}.

Very recently, $\alpha$-RuCl$_3$ became the most representative proximate Kitaev material.
Regardless of its zigzag ordered ground state in zero field, evidences of Kitaev-Heisenberg excitations (i.e. the exotic magnetic excitations associated with the prominence of Kitaev interactions) have been reported by diverse experimental tools \cite{Raman,NS-NM,NS-NP,NS-PRL,NS-S,Loidl,THz,Wellm,Richard,LeeMY,Shibauchi,MiaoH}.
More interestingly, its magnetic order can be melted by a moderate in-plane field \cite{Baek,Sears,Anja,Richard,LeeMY,WangZ}, and the sought-after half-integer quantization of the thermal Hall conductivity was claimed in a narrow field range above this critical point \cite{Kasahara}.
In the high field limit, $\alpha$-RuCl$_3$ further enters into a partially-polarized phase with a spin excitation gap which opens up linearly in field \cite{HF,Richard,Baek,Anja,WangZ}.

The rich temperature-field phase diagram of the $JK\Gamma$ model suggested by $\alpha$-RuCl$_3$ is exciting, in particular due to the putative QSL phase surrounding the quantum critical point that separates the long-range magnetic order from the partially field-polarized phase (see Fig. 1).
However, it remains to be clarified, which of the features in the phase diagram are related to the $JK\Gamma$ model, and which of them are barely material-specific properties.
The latter is a particularly complex issue since $\alpha$-RuCl$_3$ is sensitive to disorder and suffers from stacking faults and twinnings \cite{Vojta,Coldea,Yamashita}.
In this regard, it is useful if a sibling Kitaev material can be coordinatively investigated and compared to $\alpha$-RuCl$_3$.
Recently, two theoretical papers simultaneously proposed another route to realize the Kitaev model in $d^7$ cobaltates \cite{LiuHM,Motome}.
Demonstrating Kitaev interactions in the high-spin $d^7$ systems not only extends the terrain of candidate Kitaev compounds, but also introduces a mechanism to strongly reduce the $J$ interactions, thus intensifying the dominance of the $K$ term \cite{LiuHM,Motome}.
In Ref. \cite{LiuHM}, Na$_2$Co$_2$TeO$_6$ was explicitly pointed out to be such candidate.
Stimulated by these predictions, the overlooked Co-based frustrated materials again came into focus. Some surprising results have already been reported in several honeycomb and triangular Co-compounds since then \cite{ZRDd,ZRDk,ZRDt,SunXF,Trump,Stock,MaJ,Park,ChenWJ}.

\begin{figure}
\includegraphics[clip,width=0.35\textwidth]{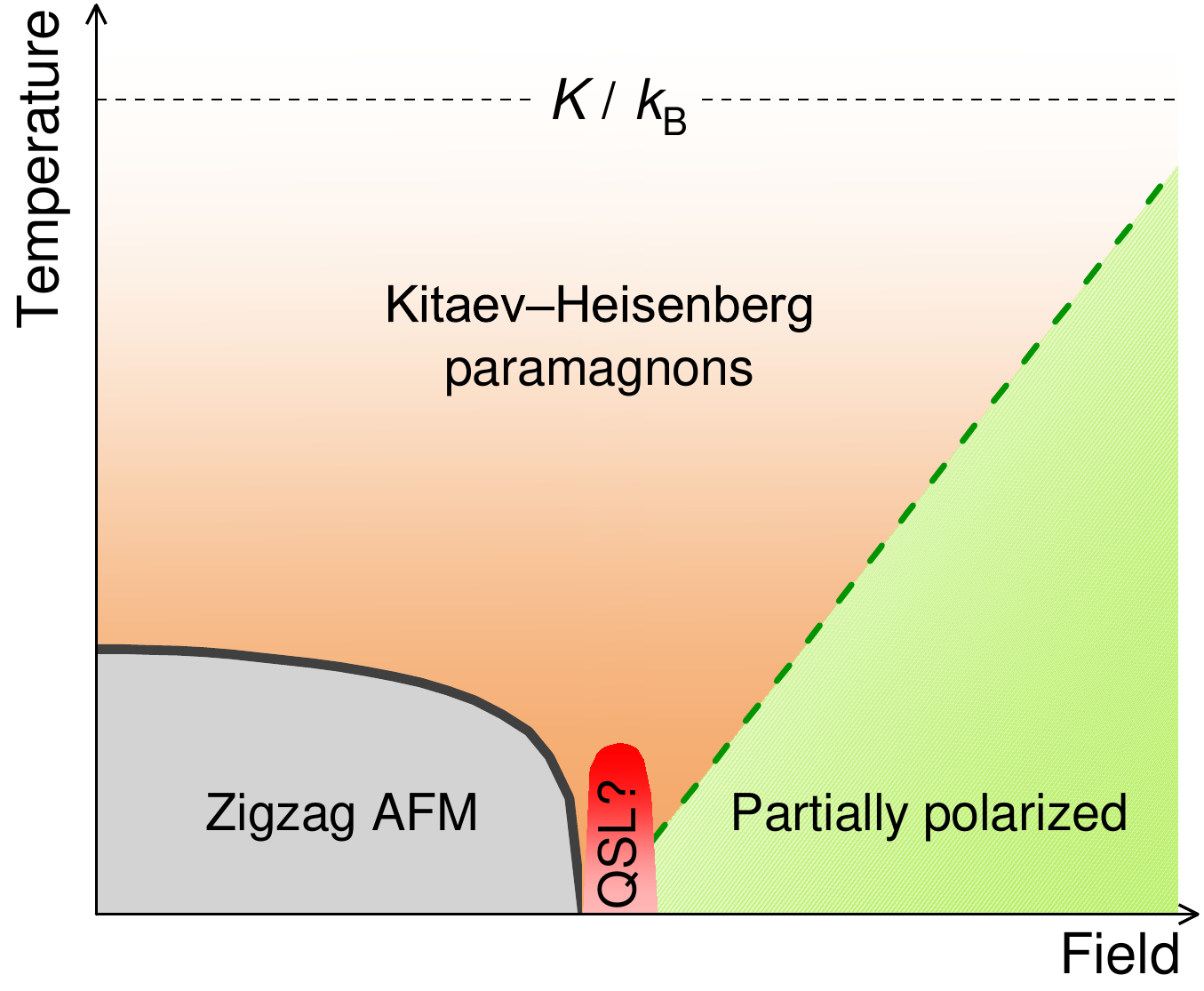}
\caption{(Color online).
Schematic sketch of the possible phase diagram of the $JK\Gamma$ pseudospin-1/2 model, shaped by the studies mainly on $\alpha$-RuCl$_3$ \cite{JPCM,TakagiRev}.
In-plane magnetic field can tune the zero-field zigzag antiferromagnetic order into the high-field (partially) polarized state with a gap which opens up linearly.
A $Z_2$ QSL state is frequently claimed to exist in a small field window at low temperature \cite{TakagiRev,NS-NP,Kasahara}.
The Kitaev-Heisenberg paramagnons are expected to survive up to high temperatures comparable to the Kitaev interaction strength \cite{NS-S,NS-NM,NS-NP,Raman,Richard,JPCM}.
}
\end{figure}

In this letter, we report the thermal conductivity $\kappa$ of Na$_2$Co$_2$TeO$_6$ single crystals from room temperature down to 6 K, with in-plane magnetic fields $B$ up to 15 T.
We unveil a strong impact of $\textbf{B}$ on $\kappa$ at low temperature: (i) beyond a magnetic field of $B_c \approx$ 10 T, $\kappa$ increases more than one order of magnitude,
and (ii) the $\kappa$($T$) curves at high fields exhibit a double-peak structure.
Remarkably, this peculiar $\kappa$($T,B$) profile is striking similar to that of $\alpha$-RuCl$_3$ \cite{Richard}.
Thus, our data strongly suggest a commonality in the physics of Na$_2$Co$_2$TeO$_6$ and $\alpha$-RuCl$_3$.
We conclude in particular, that $\kappa$($T,B$) is essentially phononic, where the predominant phonon scattering is caused by Kitaev-Heisenberg-type excitations, similar to those of $\alpha$-RuCl$_3$.

\begin{figure}
\includegraphics[clip,width=0.32\textwidth]{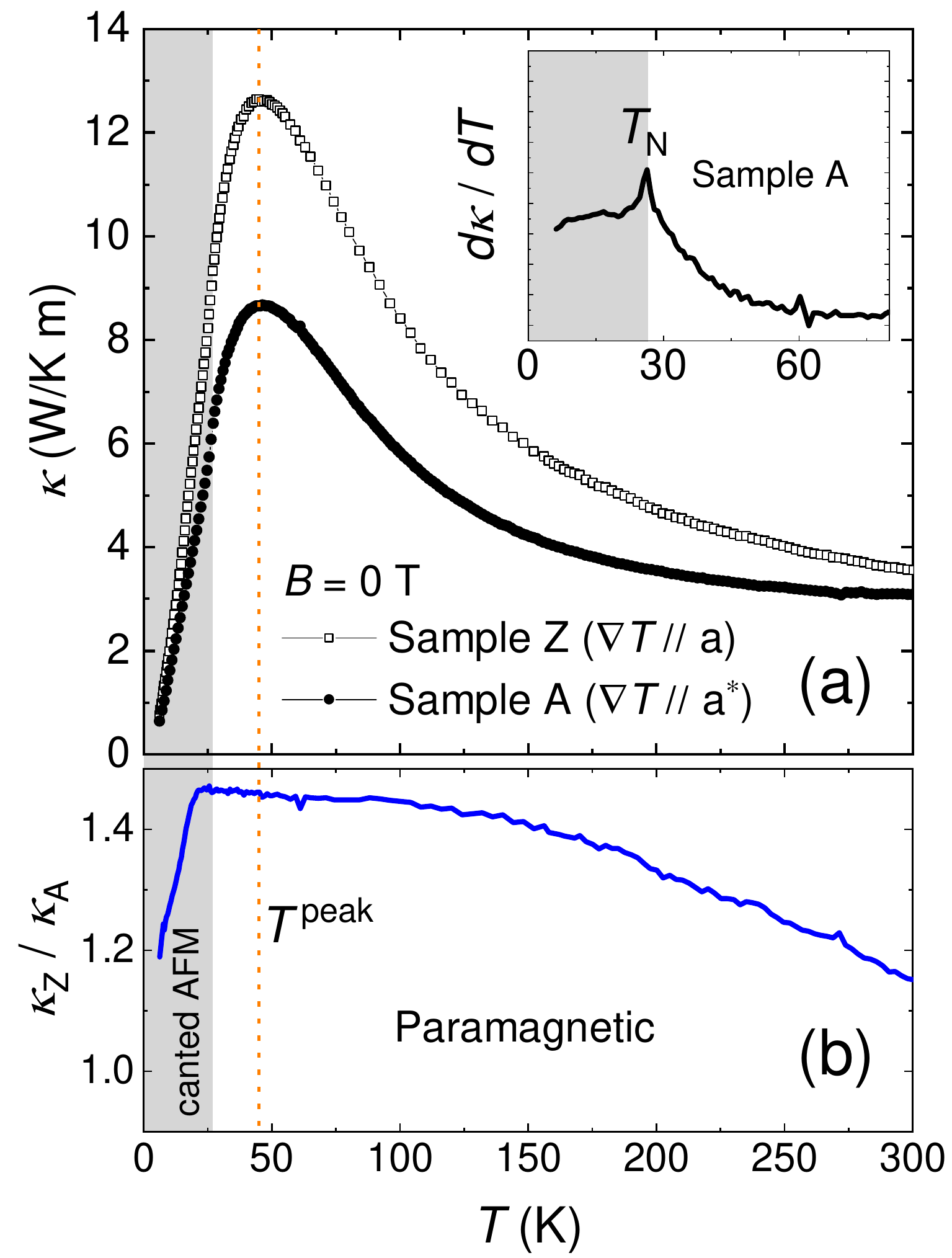}
\caption{(Color online).
(a) Temperature dependence of the thermal conductivity of two Na$_2$Co$_2$TeO$_6$ single crystals without magnetic field. The heat current was applied in two orthogonal directions, parallel to the armchair ($\kappa_A$) and zigzag ($\kappa_Z$) edges, respectively. Insert shows the derivative of the $\kappa_A$($T$) curve around the magnetic ordering temperature. (b) The ratio between $\kappa_Z$($T$) and $\kappa_A$($T$). The dashed region indicates the three-dimensional magnetically ordered state \cite{ChenWJ}. The dotted line highlights the peak temperature $T^{peak}$.
}
\end{figure}

\begin{figure*}
\includegraphics[clip,width=0.90\textwidth]{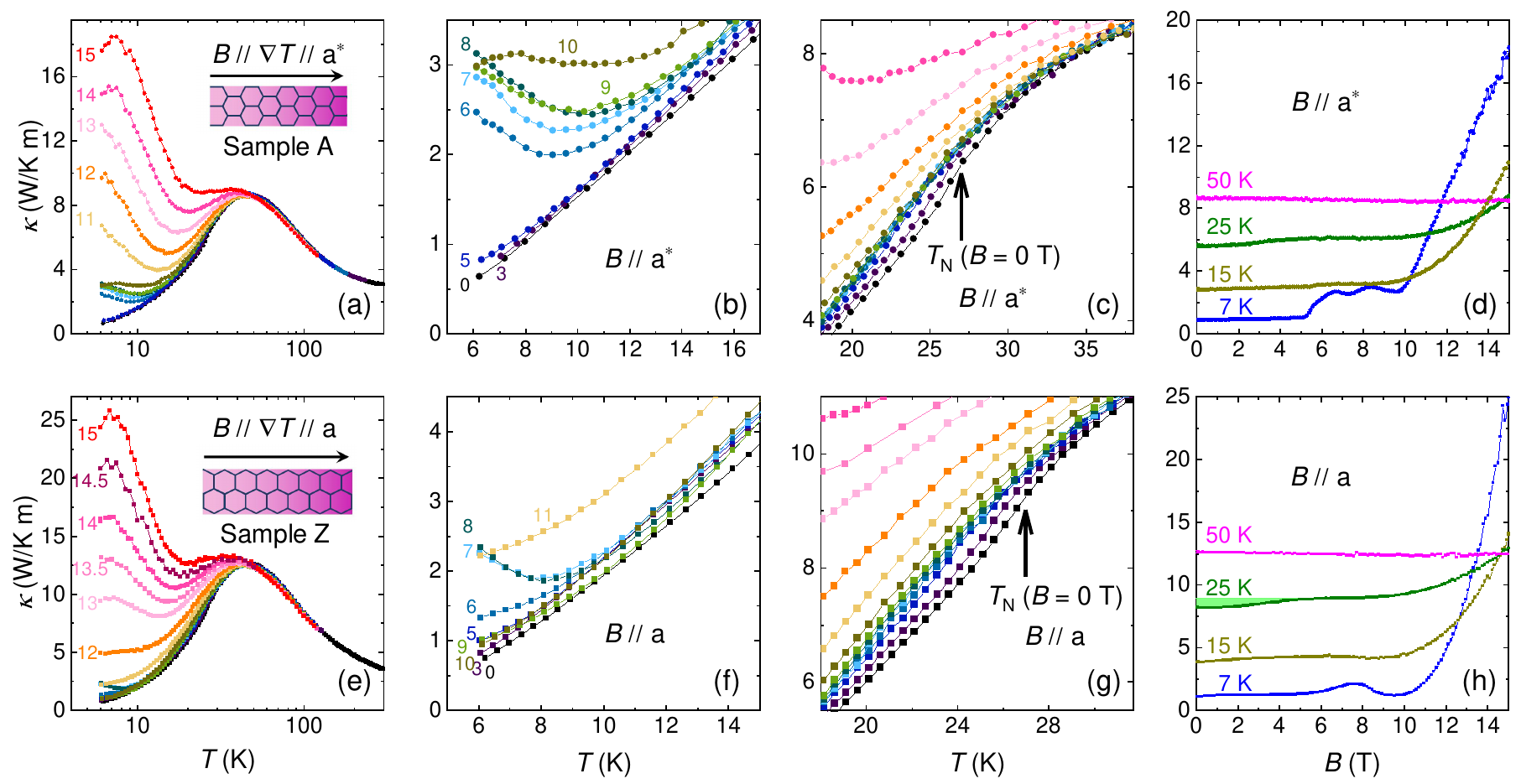}
\caption{(Color online). Thermal conductivity of Na$_2$Co$_2$TeO$_6$ crystals in various magnetic fields applied along the two in-plane high-symmetry directions. As the sketches show, the field was applied parallel to the thermal gradient. (a) Fixed-field $\kappa$($T$) curves with $\textbf{B}\parallelsum \textbf {a}^*$. The field values (in T) are indicated by the numbers next to each curve. Note that the $T$ axis is plotted in log scale for clarity.
(b) Magnified view of the low-temperature part to show the low-field curves clearly. (c) Magnified view around the magnetic transition $T_N \approx$ 26 K \cite{note}. The zero-field $T_N$ is indicated by the black arrow.
(d) $\kappa$($B$) isotherms at selected temperatures for $\textbf{B}\parallelsum \textbf {a}^*$.
(e), (f), (g) and (h) show the same aspects for sample Z with $\textbf{B}\parallelsum \textbf {a}$.
The slight increase of $\kappa$ by suppressing the magnetic order is highlighted by the light green region as an example for the 25 K isotherm in (h).}
\end{figure*}

High-quality Na$_2$Co$_2$TeO$_6$ single crystals were grown by a modified flux method \cite{Yao}.
Two regular bar-shaped samples of $5 \times 1\times 0.1$ mm$^3$ were employed in the thermal transport study. Due to the well-defined geometry, we estimate the geometric error of $\kappa$ to be less than 10\%.
The longest edges of the crystals were set parallel to the armchair ( $\parallelsum \textbf {a}^*$, Sample A) and zigzag ( $\parallelsum \textbf {a}$, Sample Z) directions of the Co-Co bonds, respectively.
Steady-state thermal conductivity measurements were performed with the standard four point geometry. One side of the sample was glued directly to the heat sink, and the other side was thermally excited by a chip heater.
The temperature gradient $\nabla T$ generated along the long edge of the sample was measured by a differential AuFe/Chromel-P thermocouple which had been calibrated carefully in magnetic field.
A custom-built high-vacuum low-noise probe provides the controlled heat sink temperature from 300 K to 6 K with a stability of about 0.1 mK.
Magnetic fields up to 15 T were generated by a commercial superconducting magnet. Fields were applied parallel to the thermal gradient.

The temperature dependent thermal conductivity $\kappa$($T$) of both samples in zero magnetic field are shown in Fig. 2(a).
At first glance, these $\kappa$($T$) curves resemble that of a conventional phononic heat conductor unaffected by novel excitations \cite{Berman}: here, $\kappa_p$($T$) $\approx C_p\nu_pl_p $, with the phononic specific heat $C_p$, velocity $\nu_p$, and mean free path $l_p$.
At low temperature, $\kappa_p$ follows the specific heat, since $\nu_p$ and $l_p$ are practically temperature independent, and thus rapidly grows with $T$.
Towards higher $T$, the reduction of the phonon mean free path $l_p$ by the sample-independent phonon Umklapp processes dictates a $1/T$-like tail of $\kappa$($T$).
A single broad peak of $\kappa_p$($T$) is thus typically present at around 1/10 or less of the Debye temperature $T^{peak} \lessapprox \Theta_D/10$ \cite{Berman}.
However, a closer inspection of the results found such trivial interpretation impeachable.
The observed $T^{peak} \approx 45$~K seems abnormally high considering $\Theta_D \approx$ 280~K which can be evaluated from the specific heat data \cite{Yao}.
Additionally, as shown in Fig. 2(b), the $\kappa_Z/\kappa_A$ ratio is essentially flat around $T^{peak}$, but drops sharply at much lower $T$, suggesting $T^{peak}$ is not the result of a canonical competition between growing $C_p$ and falling $l_p$ due to the Umklapp scattering.
Furthermore, the development of a three-dimensional long-range magnetic order (see Ref. \cite{ChenWJ}) indeed has an impact on the $\kappa$($T$) curves.
The pertinent anomaly, albeit weak, manifests itself clearly in the $T$-derivative of $\kappa$($T$) as shown in the inset of Fig.2.
For common magnets, including $\alpha$-RuCl$_3$, a sharp upturn of $\kappa$($T$) is expected below the magnetic transition temperature $T_N$, due to the suppression of the phonon-magnon scattering in the magnetically ordered state \cite{ZRDk,Richard,LeeMY}.
Notably, we find $\kappa$($T$) of Na$_2$Co$_2$TeO$_6$ to continue decreasing below $T_N$.
In our scenario of phonon-dominated heat transport which will be elaborated next, this difference may be attributed to a substantially greater ratio between $k_BT_N$ ($T_N \sim 26$ K) and the magnetic excitation gap ($\triangle \sim1$ meV) of Na$_2$Co$_2$TeO$_6$ than that of $\alpha$-RuCl$_3$ ($T_N \sim 7$ K and $\triangle \sim2$ meV) \cite{ChenWJ,NS-PRL,note}.

Our above notion of predominant phonon heat transport, which is limited at low $T$ by magnetic scattering, is further supported by the in-plane magnetic field effects on $\kappa$.
As depicted in Fig. 3, the $\kappa$($T$,$B$) profiles are qualitatively similar between the two samples under \textbf{B} applied in orthogonal in-plane directions, and also between Na$_2$Co$_2$TeO$_6$ and $\alpha$-RuCl$_3$ \cite{Richard}.
Magnetic fields have a negligible impact on the $\kappa$($T$) curves above 50 K, but changes them profoundly at lower temperatures.
The $\kappa$($T$) curves soar in high fields ($B >$ 10 T), and display a double-peak structure, resembling the peculiar features of $\alpha$-RuCl$_3$ \cite{Richard}.
The same holds for the low-temperature $\kappa$($B$) isotherms which increase dramatically above a critical field $B_c \sim$~10~T.
The strong similarity of our data to the findings for $\alpha$-RuCl$_3$ therefore corroborates our above notion of predominant phonon heat transport with a characteristic scattering due to the Kitaev-Heisenberg excitations.
These peculiar features presented by $\kappa$($T$) and $\kappa$($B$) at high fields can straightforwardly be explained by restoring the phonon conductivity via opening a gap for the Kitaev-Heisenberg excitations in the partially-polarized high-field phase, exactly the same mechanism established for explaining the mentioned features in $\alpha$-RuCl$_3$ \cite{Richard}.

\begin{figure}
\includegraphics[clip,width=0.47\textwidth]{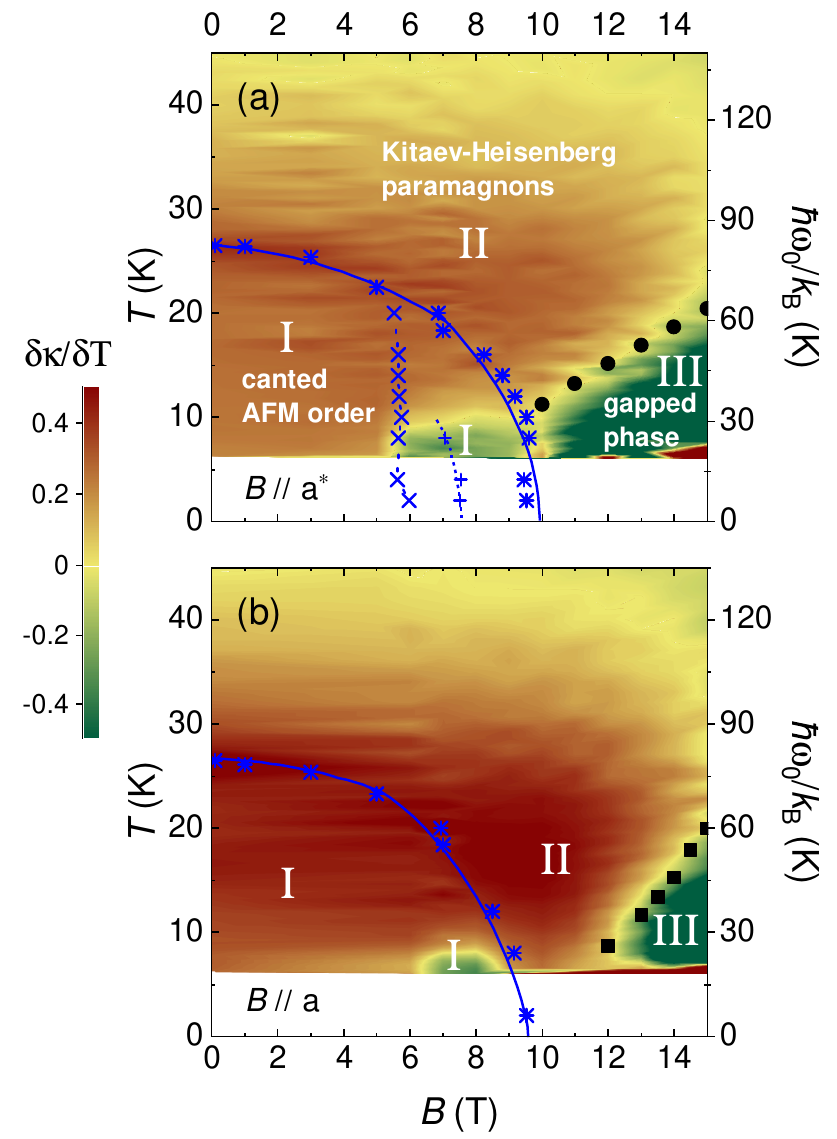}
\caption{(Color online). Color-contour representation of the $T$ derivative of $\kappa$($T,B$) with field (a) $\textbf{B}\parallelsum \textbf {a}^*$ and (b) $\textbf{B}\parallelsum \textbf {a}$. The magnetic transitions were determined by thermodynamic probes \cite{note}, and are represented by the blue points. The evolution of the onset temperature $T_N$ ($\ast$) of three-dimensional magnetic order is nearly the same for both \textbf{B} directions.
A field-induced canting reversal ($\times$) is found for $\textbf{B}\parallelsum \textbf {a}^*$ but cannot be detected for $\textbf{B}\parallelsum \textbf {a}$ \cite{Yao,note}. There is one additional transition (+) recognizable only for $\textbf{B}\parallelsum \textbf {a}^*$ at low $T$ \cite{note}. The lines are guides to the eye.
The gap energy $\hbar$$\omega_0$/$k_B$ in the partially-polarized phase extracted from the Callaway fit of the data (see text) are shown as black circles (squares) to the right ordinates.
}
\end{figure}

These results are presented in a more intuitive way in Fig. 4 as a color contour plot of $\delta\kappa$/$\delta$$T$, against the magnetic phase boundaries extracted from the magnetization measurements \cite{note}.
Three main regions (I, II and III) can be assigned by referring to the established conclusions of $\alpha$-RuCl$_3$ \cite{Richard}.
$Region$ $I$ represents the (canted) antiferromagnetic order at low-temperatures and low-fields, enclosed by the $T_N$($B$) line \cite{Yao}.
As mentioned above, the phase boundary at $T_N$ is barely visible in $\kappa$($T$) curves, and the curves bend downwards in the ordered phase. Instead of gapping out the Kitaev-Heisenberg excitations like $\alpha$-RuCl$_3$ \cite{Richard}, spin-phonon scattering is enhanced by entering this phase.
Notably, an intermediate field region labeled as $Region$ $I^*$ can be recognized below about 10~K for both \textbf{B} directions.
It is featured by an enhancement of $\kappa$, seen more clearly in the low $T$ $\kappa$($T$) curves (Fig. 3(b) and 3(f)) and the $T = 7$~K $\kappa$($B$) isotherms (Fig. 3(d) and 3(h)).
It is obvious that the more complex magnetic phases  detected via thermodynamic studies for $\textbf{B}\parallelsum \textbf {a}^*$ confer more features to the $\kappa$($B$) curve.
We are aware that an intermediate-field phase was also reported in recent $\alpha$-RuCl$_3$ studies, but it only exists for a certain \textbf{B} direction (corresponding to $\textbf{B}\parallelsum \textbf {a}$ in our representation) \cite{Balz}.
One explanation for the features of Region I$^*$ is that magnetic texture changes inside the canted AFM ordered state, which reduces the spin-phonon scattering.
Of course, it is also possible that additional transport channels of magnons are activated inside Region I$^*$. Further studies on this region are desired.

$Region$ $II$ is constituted by a narrow field range above $B_c \approx 10$~T before the system enters the higher field phase (Region III).
This Region II is smoothly connected to the broader region at higher temperatures which is dominated by the Kitaev-Heisenberg paramagnons.
It is of particular interest whether the ground state of this phase is a QSL as is conjectured for $\alpha$-RuCl$_3$ \cite{Kasahara}. Indeed, while our data for $\textbf{B}\parallelsum \textbf {a}^*$ suggest a very narrow width of a possible QSL state in Region II, the data for $\textbf{B}\parallelsum \textbf {a}$ leave a quite large range of approximately 10 $-$ 11.5~T for it.

Finally, $Region$ $III$ is the high-field phase where the $\kappa$($T$) curves acquire a double peak structure and $\kappa$($B$) exhibits a strong increase.
For $\textbf{B}\parallelsum \textbf {a}^*$, Region III seems to set in for $B > B_c \approx 10$~T, whereas for $\textbf{B}\parallelsum \textbf {a}$, our data suggest this phase to appear at $B\gtrsim 11.5$~T.
Both features imply the opening of a gap for the spin excitations in high fields, connected to the partially polarized phase \cite{HF,Richard,Baek,Anja,WangZ,MV,Winter}.
The inflection point of $\kappa$($T$), $T_{min}$, can serve as a rough gauge of the energy of the phonon-scattering magnetic modes \cite {Richard,HF}. $T_{min}$ manifests itself as the zero contour line in Fig. 4. One can also see the tendency in Fig. 3 that $T_{min}$ shifts to higher temperature in larger field.

We analyzed the high-field data based on the Callaway model as performed in $\alpha$-RuCl$_3$ \cite{Richard,note}.
The extracted gap sizes ($\hbar\omega_0/k_B$) of magnetic excitations are summarized to the right ordinates of Fig. 4.
It is clear that fields applied in the two directions open this gap roughly linearly.
The $\hbar\omega_0/k_B$ $vs$ $B$ slopes for the two field directions are slightly different, and both of them are much larger than that of $\alpha$-RuCl$_3$ \cite{Richard,Anja,Baek}.
In analogy to $\alpha$-RuCl$_3$ and according to the results of the theory and calculations for the $JK\Gamma$ model, one might speculate that $\hbar\omega_0$ can be identified with a van Hove singularity near the $\Gamma$ point \cite{Anja,Winter,MV}.
We expect that our experimentally determined parameters of Na$_2$Co$_2$TeO$_6$ will prove essential for a theoretical construction of its magnon spectrum and for finding its place in the $JK\Gamma$ parameter space.

It is interesting to apply the information gained for Region III with respect to the spin excitation gap to Region I and II, in particular at zero field.
As exhibited in Fig. 4, $T_{min}$ is related to the gap size by $\alpha T_{min}=\hbar\omega_0/k_B$ with $\alpha\approx3$ \cite{note}. Since there is no $T_{min}$ resolved in the zero field $\kappa$($T$) curves down to 6~K, we can use this information to estimate that strong spin excitations exist at least down to an energy scale of 2~meV, deep inside the magnetically ordered phase in zero field.
Besides, we stress that for $\textbf{B}\parallelsum \textbf {a}$, $\kappa$($T$) at $B\approx B_c$ is virtually identical to that at zero field down to the lowest $T$ (see Fig. 3(f)). This implies a similarly small energy scale for the spin excitations in Region II, which underpins the possible realization of a field-driven $U$(1) QSL state \cite{Hickey}.

To summarize, we presented the thermal conductivity of a new Kitaev material Na$_2$Co$_2$TeO$_6$, and explained the data as phononic heat transport strongly scattered by magnetic excitations.
In-plane magnetic field confers different ground states to Na$_2$Co$_2$TeO$_6$.
Strong scattering survives below $T_N$ in the low-field magnetically ordered state Region I.
A linearly opened excitation gap is extracted from the unusual double-peak $\kappa$($T$) curves in the high-field partially polarized state Region III.
In between them is the most exciting state Region II (and perhaps also Region I$^*$), where strong low-energy spin fluctuations are present despite the absence of magnetic order. Especially for $\textbf{B}\parallelsum \textbf {a}$, we found a relatively wide field window suitable for a QSL state.
Overall, our results support the conjecture that Na$_2$Co$_2$TeO$_6$ is another materialization of the Kitaev model.
All the interesting results of $\alpha$-RuCl$_3$ deserve seeking for in Na$_2$Co$_2$TeO$_6$, that will promote the understanding of proximate Kitaev materials.

We would like to thank Vladislav Kataev, Christoph Wellm, and Xenophon Zotos for fruitful discussions, and thank Juliane Scheiter for technical support.
This work has been supported by the Deutsche Forschungsgemeinschaft through SFB 1143 (project-id 247310070), the W\"{u}rzburg-Dresden Cluster of Excellence on Complexity and Topology in Quantum Matter-ct.$qmat$ (EXC 2147, Project No. 390858490). This work has further been supported by the European Research Council (ERC) under the European Union's Horizon 2020 research and innovation programme (Grant Agreement No. 647276-MARS-ERC-2014-CoG).
Work at Peking University was supported by the NSF of China under Grant No. 11888101, and by the NBRP of China under Grant No. 2018YFA0305602.

\end{document}